\documentclass{article}
\usepackage[utf8]{inputenc}
\usepackage{geometry}      
\usepackage{times}
\usepackage{amsmath,amssymb,amsthm,graphicx,color,bm}
\usepackage{enumerate}

  \newcommand{\da}{\dagger}

  \newcommand{\rh}{\rho}

  \newcommand{\cH}{{\mathcal H}}
  \newcommand{\cI}{{\mathcal I}}
 \newcommand{\cK}{{\mathcal K}}  

\title{Echoing the recent Google success: Foundational Roots of  Quantum Supremacy}                     
\author{Andrei Khrennikov}

\begin{document}
\maketitle


\begin{abstract} The recent Google's claim on breakthrough in quantum computing is a gong signal for further  analysis of 
foundational  roots of (possible) superiority of some quantum algorithms over the corresponding classical algorithms.
This note is a step in this direction. We start with critical analysis of rather common reference to entanglement and quantum 
nonlocality as the basic sources of quantum superiority.  We elevate the role of the Bohr's {\it principle of complementarity}\footnote{} (PCOM) by interpreting the Bell-experiments as statistical tests of this principle. (Our analysis also  includes comparison of classical vs genuine quantum entanglements.) After a brief presentation of PCOM and endowing it with the information interpretation, we analyze its computational counterpart. The main implication of PCOM is that by using the quantum representation of probability, one need not compute the joint probability distribution (jpd) for observables involved in the process of computation. Jpd's calculation is  exponentially  time consuming. Consequently, classical probabilistic algorithms involving calculation of jpd for $n$ 
random variables can be over-performed by quantum algorithms (for big values of $n).$ Quantum algorithms are based on quantum probability calculus. It is crucial that the latter modifies the classical formula of total probability (FTP). Probability inference 
based on the quantum version of FTP leads to constructive interference of probabilities increasing probabilities of some events. 
We also stress the role the basic feature of the genuine quantum superposition comparing with  the classical wave superposition:  generation of discrete events in measurements on superposition states. Finally, the problem of superiority of quantum computations is coupled with the quantum measurement problem and linearity of dynamics of the quantum state update. 
\end{abstract}

\section{Introduction}

The recent tremendous success in engineering of quantum computers generated the new wave of interest to foundational analysis of their functioning. This interest is heated by widely distributed in media claims that they beat classical digital computers as well 
as intensive critique of such claims - especially the recent Google-team claim \cite{Google,Google1} (and its questioning in 
\cite{IBM,Karl}). Such type of controversy cannot be resolved  just through pumping money and news. 

Unfortunately, {\it the foundational grounds of quantum computing are really shaky.}\footnote{We mean really foundational grounds, not 
formal mathematical propositions.}   
The possibility to generate {\it superpositions} and operate with them  is often pointed as one of the distinguishing features of quantum computing. However, classical wave devices can also  
generate superpositions,  e.g.,  {\it optical computers.} Typically states' {\it entanglement} is considered as the crucial component of quantum computing. However,  entanglement is just a mathematical structure and its physical meaning is still 
the subject of intensive foundational debates related to such an ambiguous notion as {\it quantum nonlocality} 
(``action at a distance''). The main problem is that quantum computational studies and engineering practically ignore this interpretational challenge. 
People just repeatably  point to entanglement  as the basic source of superiority of quantum computing.\footnote{ However, no matter how much you say the sugar in your mouth will not become sweeter. The same with entanglement: no matter how often you say ``entanglement'', your understanding of  quantum computing would not become better.} 

We can point to two main tendencies in researchers' behavior:
\begin{enumerate}
\item To be totally satisfied by formal mathematical representation of entanglement as state's non-separability.
\item To sanctify entanglement (and quantum computing) by referring to  quantum nonlocality.
\end{enumerate}

To avoid involvement into deep foundational discussions, typically researchers working on quantum computing theory use the first strategy and the essential part of this theory is simply linear algebra in complex Hilbert space. One of our aims  is to show that the formal mathematical treatment of entanglement cannot justify supremacy of quantum computing. Following recent paper \cite{KHR1}, we  criticize  common referring to quantum nonlocality as the physical basis of entanglement. We elevate the role of the {\it Bohr's principle of complementarity} \cite{BR0,PL2} (PCOM) by interpreting the Bell-experiments as statistical tests of this principle (section \ref{ENT}).  

Our analysis also  includes comparison of classical vs genuine quantum entanglements (section \ref{ENT}) implying comparison of classical optical vs genune quantum computing. As we know (see, e.g., \cite{CE1}-\cite{CE5}), the classical electromagnetic field can be successfully used to model some basic features of genuine quantum physical systems. In particular,  such classical field modeling is a helpful tool for  simulation in quantum information theory. It can be easily shown that from the formal mathematical viewpoint entanglement of a few degrees of freedom of classical electromagnetic field does not differ from entanglement of genuine quantum states. We discuss consequences for quantum computing of this mathematical classical-quantum isomorphism. At the same time, we criticize (see also \cite{KHR3}) so common in classical optics prejudice that classical and quantum entanglements differ due to 
quantum nonlocality (e.g., \cite{CE1,CE2}).

After excluding foundational mysticism from quantum computing (such as instantaneous state update via action at a distance), we 
start analysis of (quantum physical) foundations of quantum computing based on PCOM by starting    
with its  brief presentation of PCOM (section \ref{LL}). The main implication of PCOM (section \ref{COMPUT}) is that by using the quantum representation of probability, one need not compute the {\it joint probability distribution} (jpd) for observables involved in the process of computation. Jpd's calculation is exponentially time consuming. Consequently, {\it classical probabilistic algorithms involving calculation of jpd for $n$  random variables can be over-performed by quantum algorithms} (for big values of $n).$ We stress that our analysis is about comparison of  concrete  classical vs. quantum algorithms. (Even if straightforward classical digital algorithmization leads to jpd-calculation, some additional constrains can simplify classical computations, e.g., with the aid of Bayesian nets.)

 Quantum algorithms are based on quantum probability theory (section \ref{QP}). The quantum probability calculus modifies the classical formula of total probability (FTP). Probability inference  based on its quantum version (FTP with the interference term \cite{KHR4}, section \ref{FTP}) can lead to increase of probabilities of some events (via their constructive interference). It is important to stress that this is interference of probabilities, not waves. We point the basic feature of the genuine quantum superposition comparing with  the classical wave superposition:  generation of discrete events in measurements on superposition states 
(Bohr's quantum phenomena \cite{BR0,PL2,KHR3}).  

The problem of supremacy of quantum computations is coupled with the quantum measurement problem and linearity of dynamics of the quantum state update in the process of measurement (section \ref{PP}). We also emphasize the role of the frequency realization 
of quantum probabilities in coupling with measurement's role (section \ref{FR}). 

The most interesting question is about the role of the existence of the indivisible quantum of action (the Planck constant) 
in quantum computational supremacy. We discussed it section \ref{LL} in connection with Bohr's formulation of PCOM. Typically, 
this question is ignored. (Generally the role of $h$ in quantum computing is not so much highlighted.)

\section{Classical versus quantum superposition}
\label{SUP}

The bit is a portmanteau of binary digit and this preassumes the discrete
structure of information represented by bits. In Wikipedia, it is stated that

\medskip

{\it ``Qubit is the quantum version of the classical binary bit that is physically
realized with a two-state device. A qubit is a two-state (or two-level)
quantum-mechanical system, one of the simplest quantum systems displaying
the peculiarity of quantum mechanics.''} 

\medskip

Although Wikipedia is not the
best source of citation in a research article, precisely this definition is 
widely used by researchers working in quantum information theory. 
(The fundamental notion of qubit was invented in Schumacher's paper  \cite{Schumacher}, where  qubit was introduced by ``replacing the classical idea of a binary digit with a quantum two state system.'')  The second part of Wiki's text reflects the common neglect by the role of measurement and repeats Schumacher's message. At the same time, the first part points perfectly to a two-state
device, the source of discrete counts; a device that can distinguish two states.
Unfortunately, in quantum information research one typically operates with
states forgetting about extracting information from them. Consequently, the essential part of this research is reduced 
to linear algebra in complex Hilbert spaces. This state based activity is operationally convenient and it played the important 
role in rapid development of quantum information theory. However, foundationally it is a dangerous, since the role of the difference between the classical wave and genuine quantum superpositions is not emphasized. This leads to some casualties in realization of the quantum information project. One of the basic foundational problems was generated by extensive studies of the classical optics analogs 
of the quantum information structures \cite{CE1}-\cite{CE5}. However, in these studies the aforementioned difference of classical and quantum superposition was shadowed by the emphasis of  the difference between classical and quantum entanglements. 
And I claim that the main problem is superposition, not entanglement  (see \cite{KHR3}).
The first and best comparison of  quantum and classical superposition is presented in Dirac's book \cite{Dirac}, p.13:

\medskip

{\it ``The nonclassical nature of the superposition process is brough out clearly if we consider the superposition of two states
$A$ and $B,$ such that there exists an observation which, when made on the system in state $A,$ is certain to lead to one particular 
result, $a$ say, and when made on the system in state $B,$ is  is certain to lead to some different
result, $b$ say. What will be the result of the observation when made on the system in  the superposition state? The answer is that the result will be sometimes $a$ and sometimes $b,$ depending on the relative weights of $A$ and $B.$''}

\medskip

Here Dirac clearly point to crucial role of an observation (see also Jaeger \cite{Jaeger2}, p.1) and the characteristic feature of the quantum observation, namely, 
its ability to produce ``one particular result.'' This discreteness of quantum observations, e.g., clicks of photodetectors, or dots 
on a screen covered by photoemulsion, was emphasized by Bohr who formalized this feature in the notion of phenomenon  
\cite{BR0,PL2,KHR3} (see section \ref{LL}) for further discussion). 

{\it  The possibility to extract from the signal of an individual value (to observe phenomenon) is one of the basic features of quantum information theory and it plays the crucial role in quantum computing.} 

The superposition is the crucial condition of realization of quantum computing, but it has to be understand  in the right way as the genuine quantum superposition (see again Dirac \cite{Dirac}, see also \cite{KHR3}). The notion of qubit as solely state characterization is misleading. We shall continue to discuss the  role of the quantum superposition  by coupling 
it with Bohr's PCOM.

\section{Entanglement and quantum nonlocality or complementarity of quantum observables?}
\label{ENT}

We turn to aforementioned two behavioral tendencies in handling of entanglement, see introduction. 

Although the formal mathematical treatment of entanglement is convenient (especially in quantum engineering), it 
has its own problems. One of such problems became visible as the result of  intensive 
studies on so-called {\it classical entanglement} (see, e.g., for reviews \cite{CE1}-\cite{CE5}). It can be  easily shown that the mathematical structure 
for a few degrees of freedom of the classical electromagnetic field is identical to  the mathematical structure of a compound genuine quantum system. Entangled (in the mathematical sense) states can be generated by beams of classical  light. This means that the strategy based on the formal mathematical treatment of entanglement and complete ignoring the interpretational problems was not successful. It is impossible to proceed without trying to find a consistent physical interpretation fo entanglement.   And commonly the second behavioral strategy is in use: quantum systems (such as electrons or photons) are so unusual that they can have really mystical features, e.g., such as action at a distance - quantum nonlocality. Even gurus of classical entanglement claim \cite{CE1}-\cite{CE4} that quantum nonlocality 
is basis source of  ``true quantum entanglement'' of states of subsystems of a compound genuine quantum system (cf., however, 
with latter review \cite{CE5}). They sharply distinguish  it from ``fake entanglement'' of degrees of freedom  of a single system, either quantum or classical. These two physical realizations of mathematically defined entanglement are known as {\it inter and intra system} entanglements.  

However, the notion of  quantum nonlocality is ambiguous and history of its invention is really tricky 
(see, e.g., \cite{KHR3}). This notion is a mixture of quantum and subquantum considerations culminating in theoretical and experimental studies on the Bell type inequalities. In discussions on quantum nonlocality, people freely mix nonlocality of genuine quantum theory with possible nonlocality of hypothetical subquantum models. The first step to clarification should be separation of two types of nonlocality. And this step was recently done in author's paper \cite{KHR1}. It was shown that {\it local incompatibility of observables}, i.e., observables $A_j, j=1,2,$ of Alice (and $B_j, j=1,2,$ of Bob) is necessary 
and sufficient  for violation of the CHSH-inequality (for at least one state). 
Mathematically local incompatibility is described as $[A_1,A_2] \not=0$ and $[B_1,B_2] \not=0.$  
Thus, {\it  Bell's type inequalities are just special statistical tests of the Bohr's complementarity principle}  \cite{KHR1}.  

\medskip

{\it Quantum mechanics is  a local theory (constrained by PCOM).}

\medskip

At the same time, some subquantum theories can have  nonlocal features, e.g., Bohmian mechanics. Such subquantum 
nonlocality can be explored for clarification of distinguishing features of quantum computing. This is an interesting foundational project of big complexity. One of the difficulties is that the meaning of ``subquantum'' is fuzzy. We can imagine a  huge variety of couplings between
various classical physical models and the quantum mechanical model. We can mention the von Neumann \cite{VN} and Bell \cite{Bell} theorems as two the most known examples of such subquatum$\to$quantum couplings. Bell by himself noted that may be 
 ``no-go theorems'' including his own theorem are just signs of luck of imagination \cite{Bell}. 
However, human imagination is so powerful that without direct coupling to physics it can generate a variety of subquatum models providing  exotic justifications of superiority of quantum computers. 
In this paper we shall not consider subquantum models and proceed in the purely quantum mechanical framework. 

\medskip

In the light of the critical argumentation given in article  \cite{KHR1},  the quantum nonlocality interpretation of entanglement lost its power.  The following natural question arises:

\medskip

{\it Can superiority of quantum computing be justified by PCOM?}
 
\section{Complementarity principle}
\label{LL}

Typically PCOM is considered as difficult for understanding. I try to explain its meaning 
(see also \cite{BR0,PL2}). In 1949,  Bohr \cite{BR0} presented the essence of complementarity in the following widely citing statement:

\medskip

{\it “This crucial point ...  implies the impossibility of any sharp separation between the behaviour of atomic objects and the interaction with the measuring instruments which serve to define the conditions under which the phenomena appear. In fact, the individuality of the typical quantum effects finds its proper expression in the circumstance that any attempt of subdividing the phenomena will demand a change in the experimental arrangement introducing new possibilities of interaction between objects and measuring instruments which in principle cannot be controlled. Consequently, evidence obtained under different experimental conditions cannot be comprehended within a single picture, but must be regarded as complementary in the sense that only the totality of the phenomena exhausts the possible information about the objects.”} 

\medskip

By PCOM observables do not deliver the genuine 
properties of   quantum systems. Their outputs  result from the interaction of a system $S$ and 
a measurement instrument $M.$  The impossibility of separation of features of $S$ from interaction with $M$ implies the existence of complementary complexes of experimental conditions (experimental contexts). Logically there is no reason to assume that information obtained for different contexts can be always consistently comprehended within a single (theoretical) picture. Moreover, we have to take with surprise  the theories in that such comprehension is always possible. From this viewpoint,
non-complementarity of observables in classical physics is more surprising than complementarity of some observables in quantum physics.  

Only this part of PCOM is well known, typically under the name of {\it the wave-particle duality.} 
(We point that Bohr by himself had never used the latter notion.) We remark that, for Bohr, existence of complementary observables is just a natural consequence of the inseparability part of PCOM (which is typically ignored). Moreover, complementarity is often treated as
physical impossibility of joint measurements, say it is impossible to measure jointly position and momentum of a quantum system. 
However, from the citation it is clear that, although Bohr did not write about information, his statement is informational per its nature. 
In section \ref{COMPUT}, we shall discuss this issue in connection with quantum computing.

PCOM  has another part that even less discussable than its inseparability part. It is also encoded in Bohr's statement, as {\it ``the individuality of the typical quantum effects''}. What does it mean this ``individuality''? Again in my personal vision, this is discreteness of outputs of quantum observables: individuality of the act of measurement, say the click of a photo-detector with the possibility to couple this click with the individual photon. In fact, Bohr paid a lot of attention to this issue. Individuality was formalized in the notion of  {\it phenomenon}\footnote{Unfortunately, this part of Bohr's foundational study is practically unknown and it is discussed only by philosophers working on quantum foundations \cite{Jaeger1,PL2}.}:

\medskip

{\it `` ... in actual experiments, all observations are expressed by
unambiguous statements referring, for instance, to the registration of the point at which
an electron arrives at a photographic plate. ...  the appropriate physical interpretation of the symbolic quantum mechanical
formalism amounts only to predictions, of determinate or statistical character,
pertaining to individual phenomena ... .} (\cite{BR0}, v. 2, p. 64)

\medskip

Thus, although quantum theory produces statistical predictions, its observables generate individual 
phenomena.  Discreteness of detection events is the fundamental feature of quantum physics 
justifying existence of quantum systems, carriers of quanta.  In the rigorous quantum theoretical framework, this discreteness is formalized as non-coincidence detection for say photons emitted by a single photon source and passing a beam splitter. This feature of quantum measurements can be probabilistically encoded in the coefficient of second order coherence, $g^{(2)}(0) <1$ (see 
\cite{GR,GRT,KHR3}). 

In paper \cite{KHR3}, the difference between  ``classical entanglement'' and ``true quantum entanglement'' was analyzed. This analysis elevated the role of {\it quantum individuality} (discrete outputs of quantum measurement) as the 
alternative to quantum nonlocality in understanding the difference between classical and quantum entanglements. 

Now, we discuss the last part of Bohr's formulation of PCOM, ``... the totality of the phenomena exhausts
the possible information about the objects.'' I understood is as follows: by using complementary experimental contexts, 
we are able to collect the complete information about features of a quantum system encoded in its state. From this viewpoint,
quantum mechanics is a complete theory (cf. with the EPR-Bohr debate \cite{EPR,BR}).  

We remark that, in fact, PCOM contains another part that is not present in the above citation (but see \cite{BR0,KHR2}). This is referring to the existence of indivisible quantum of action given by the Planck constant $h$ (see \cite{} for discussion). According to Bohr, the 
limit in the magnitude of action implies the impossibility to separate the genuine properties of quantum system $S$ from the impact of the interaction with measurement device $M.$ So, the Planck constant is the root of PCOM. (At the same time its concrete value seems to be unimportant.) 

Thus, PCOM can be presented as following five interconnected statements (see also \cite{KHR2}):
\begin{itemize}
\item {\bf PCOM 1} Planck constant: Existence of the indivisible quantum of action.
\item {\bf PCOM 2} Inseparability: Dependence of measurement's output on the experimental context.
\item {\bf PCOM 3} Complementarity: Existence of complementary experimental contexts. 
\item {\bf PCOM 4} Individuality: Discreteness of quantum measurements.
\item {\bf PCOM 5} Completeness: Complementary observations  provide complete information 
about system's state. 
\end{itemize}

We remark that according to Bohr, ${\bf PCOM} 1 \implies {\bf PCOM} 2  \implies  {\bf PCOM} 3$; on the other hand, {\bf PCOM 4} and {\bf PCOM 5} are independent principles.    

Regarding {\bf PCOM 1}, it is not clear whether the existence in nature of  irreducible quantum of action  (the Planck constant) is relevant to computing and quantum information generally. We would like to underline this problem:

\medskip

{\it What is the role of Planck constant $h$ in supremacy  of quantum computing?} 
  
\medskip

I do not have an answer to this question. It seems that in the formal quantum information theory $h$ is not present at all,
The Planck constant is neither present in  common reasoning on quantum nonlocality based on the Bell inequality.\footnote{Once at one 
of the V\"axj\"o conferences on quantum foundations, one of the world's leading experimenter in quantum optics and foundations gave a talk on the final loophole free experiment on violation of a Bell type inequality. And one old physicist was very disappointed: ``You speak about quantum physics. But, where is the Planck constant?'' He got the very interesting and unexpected answer: ``There are two 
parts of quantum physics: one is about $h$ and another is without $h.$''}

\section{Superposition principle as linear representation of complementarity principle}
\label{LIN}

We remark that  quantum complementarity is a very special version of complementarity (as the general measurement principle). 
PCOM is based on the complex Hilbert state space, representation of observables by Hermitian 
operators, and noncommutativity of operators as the mathematical expression of incompatibility of observables.   
And the superposition principle (Dirac \cite{Dirac}) can be considered as the linear space representation of PCOM (Bohr \cite{BR}). (Once again, we repeat that the superposition of two states has to be understand quantumly, i.e., in coupling with quantum phenomena - individual outputs of observations.)

Thus, basically application of PCOM to quantum information theory is in the form of the possibility to expand a
quantum state vector with respect to various orthogonal bases. Physically each basis  represents 
experimental conditions (context), mathematically it represents a Hermitian operator. This operator is symbolic representation of 
a quantum measurement process generating individual (discrete) outputs. 

Linear space representation is one of the crucial necessary conditions of quantum computing. However, it is not the sufficient condition to approach its supremacy. As we have seen, classical wave systems also have linear representation. The crucial point is the possibility 
to extract discrete information from a state. 

As we shall see in section, linear state representation can generate essential increase of probability through linear interference, but 
this probability by itself is not useful for computing. Only the possibility of its {\it frequency realization}, via a sequence of discrete events makes interference of the practical use. 

\section{Quantum supremacy from quantum probability calculus}

\subsection{Complementarity and quantum probability}
\label{COMPUT}

We now analyze  {\bf PCOM 3} (based on {\bf PCOM 1} and {\bf PCOM 2}) in more detail. We proceed with its information interpretation
that is the most appropriate for quantum computing (see  \cite{V1} and especially \cite{PL1}). As we know, information is firmly coupled to probability. Hence, {\bf PCOM 3} is about the impossibility of 
construction of the comprehensive  probabilistic description of data collected through measurements of a few arbitrary quantum 
observables. Observables for which such description is possible are called {\it compatible} and those for which it is impossible are called  {\it incompatible.} However, the meaning of ``comprehensive probabilistic description'' cannot be specified without specification of a mathematical model of probability. And  it seems to be natural to follow Kolmogorov \cite{K} (1933) and consider his measure-theoretical model of probability.    

\subsection{Classical probability}
\label{CP}

We are interested in the  principles of Kolmogorov probability theory  (known as the classical probability theory)  and how violations of some of them can lead to 
supremacy of quantum computing. The {\it Kolmogorov probability space}\index{Kolmogorov probability space} \cite{K} is any triple
$$(\Lambda, {\cal F}, P),$$ 
where $\Lambda$ is a set of any origin and ${\cal F}$ is a
$\sigma$-algebra of its subsets\footnote{A collection of sets that is closed with respect to operations of countable 
union, intersection, and complement (or in logical terms: disjunction, conjunction, and negation).}, $P$ is a probability measure on ${\cal F}.$
The set $\Lambda$ represents random parameters of the model.
Kolmogorov called elements of $\Lambda$ {\it elementary events.} This terminology is standard in mathematical literature. Sets of elementary events belonging to  ${\cal F}$ are regarded as {\it events}.

Probability measure is additive, i.e.
\begin{equation}
\label{AD}
p(E_1 \vee E_2) =   p(E_1) +  p(E_2), 
\end{equation}
for disjoint events $E_j \in {\cal F},$ $E_1 \wedge E_2= \emptyset.$\footnote{In his axiomatic, Kolmogorov used generalization of additivity for 
countable sequence of disjoint events. However, he by himself pointed that this is the purely mathematical constraint. Physically
countable additiviy is not testable \cite{K}.} Another  basic feature of the classical probabilistic model is the definition of the conditional probability by the {\it Bayes' formula}\footnote{We point to the common prejudice that  the  Bayes' formula is a theorem. Not, at all! This is the classical way to introduce conditioning.}:
\begin{equation}
\label{AD1}
p(B\vert A) \equiv p(B \wedge A)/P(A),\;  P(A)> 0.
\end{equation}

In classical probability, for any group 
of random variables $A_1,...,A_n$ (representing observables), their {\it joint probability distribution} (jpd) is well defined,  
$$
p_{A_1 ...A_n}(x_1,...,x_n)=  P (\lambda \in \Lambda: A_1(\lambda) =x_1, ..., A_n(\lambda) =x_n).
$$ 
Moreover, to operate consistently, we have to compute jpds for all possible combinations 
of observables. They can be found as marginals of $p_{A_1 ...A_n},$ The classical version of  {\bf PCOM 5} is
that probability measure $P$ (which can be treated as the jpd of all random variables)  represents the complete state of a system. 

\subsection{Quantum probability}
\label{QP}

Now, we briefly recall the basics of quantum calculus of probability (see, e.g., \cite{PR77}).
Consider a pure quantum state $\psi$ and quantum observables $A$ and $B$ 
with discrete spectra. For simplicity, we consider representation of observables by
 Hermitian operators with nondgenerate spectra; $(\vert \alpha_i\rangle)$ and $(\vert \beta_i\rangle)$ denote
their eigenbases.  We recall that quantum probability is defined by Born's rule:
\begin{equation}
\label{ADz}
q(B=\beta_i)\equiv q(B=\beta_i; \psi)= \vert \langle \psi\vert \beta_i\rangle\vert^2, \; 
q(A=\alpha_i)\equiv q(A=\alpha_i; \psi)= \vert \langle \psi\vert \alpha_i\rangle\vert^2.
\end{equation}
Quantum conditional probability $q(B=\beta_i \vert A=\alpha_j)$ is defined as the probability w.r.t. the post-measurement state  
corresponding to the concrete output $A=\alpha_j.$ In the case of non-degenerate spectrum of $A,$ the post-measurement state is simply 
eigenvector $\vert \alpha_j\rangle$ (see section \ref for the general case):
\begin{equation}
\label{ADy}
q(B=\beta_i \vert A=\alpha_j) = \vert \langle \beta_i \vert \alpha_j \rangle\vert^2.
\end{equation}
In the case under consideration, non-degenerate spectra, conditional probability is symmetric, i.e., 
$q(B=\beta_i \vert A=\alpha_j)= q(A=\alpha_j\vert B=\beta_i).$ However, for non-degenerate spectra, symmetry 
of conditional probabilities is violated. So, calculation of 
quantum conditional probability $q(B=\beta_i \vert A=\alpha_j)$ consists of two steps: 
\begin{itemize}
\item {\bf Proj} Projection of the initial state $\psi$ onto eigenstate $\vert \alpha_j\rangle,$ as the result of the $A$-measurement.
\item {\bf Prob} Calculation of the probability of $B$-output w.r.t. $\vert \alpha_j\rangle.$    
\end{itemize}

\medskip

{\it  The impossibility to define jpds (see {\bf PCOM 3}) for incompatible quantum observables led to discovery of the novel probability calculus, quantum probability theory.} Here, instead of manipulating with jpds, one uses quantum state vectors and 
their transformations.

The algorithmic  power of the impact quantum probability calculus is also coupled to {\bf PCOM 5}. By operating with incompatible observables, one can extract {\it complete information} about the state of a quantum system.\footnote{We recall that in classical probability 
calculus complete information is encoded in jpd.} Moreover, by {\bf PCOM 4} this information can be represented in the form of discrete
events, phenomena.     The latter is crucial for the frequency realization of probability (see sections \ref{FR}, \ref{PP}).

Of course, the quantum probability theory was not just a mathematical discovery of a nonclassical probability model (similar to the discovery by Lobachevsky of the first  model of non-Euclidean geometry). The crucial point is  that creation  of the quantum probability theory was based on the  discovery of a wide class of physical systems following the laws of this theory. The most important feature of these systems is represented by {\bf PCOM 5}. Complete information about the state of a quantum system can be extracted from 
separate measurements of quantum observables. In contrast to classical systems, for quantum systems there is no need in joint measurements.    

This is the good place to cite Feynman  \cite{FeynmanP}  (italic was added by the author of this paper):

``From about the beginning of the twentieth century experimental physics
amassed an impressive array of strange phenomena which demonstrated the
inadequacy of classical physics. The attempts to discover a theoretical structure
for the new phenomena led at first to a confusion in which it appeared
that light,and electrons, sometimes behaved like waves and sometimes like
particles. This apparent inconsistency was completely resolved in 1926 and
1927 in the theory called quantum mechanics. The new theory asserts that
there are experiments for which the exact outcome is fundamentally unpredictable,
and that in these cases one has to be satisfied with computing
probabilities of various outcomes.   {\it But far more fundamental was the discovery
that in nature the laws of combining probabilities were not those of the
classical probability theory of Laplace\footnote{It seems that Feynman was unaware about the mathematical 
formalization of probability by Kolmogorov \cite{K}; so he referred to Laplace, not to Kolmogorov.}.}''

\section{Interference of probabilities as the basis of quantum superiority}
\label{FTP}

Consider the classical model of probability. Its crucial property is that all observables 
(described by it) can be represented as random variables with respect to the same probability measure $P.$ 
By using formulas (\ref{AD}), (\ref{AD1}),  we can easily derive the formula of total probability (FTP). Consider two discrete random variables 
$A=\alpha_1,.., \alpha_n$ and $B= \beta_1,...,   \beta_n.$ Then 
\begin{equation}
\label{AD2}
p(B=\beta_i) =\sum_j p(A=\alpha_j) p(B=\beta_i \vert A=\alpha_j) 
\end{equation}

However,  in quantum probability calculus FTP
is violated (see, e.g., \cite{KHR4,PR77}). It is transformed into FTP with the interference term (a perturbation of the 
classical formula).  By expanding $\psi$ with respect to orthonormal bases consisting of eigenvectors 
of $A$ and $B,$  we obtain \cite{KHR4,PR77}: 
\begin{equation}
\label{AD3}
q(B=\beta_i; \psi) =\sum_j q(A=\alpha_j; \psi) q(B=\beta_i \vert A=\alpha_j) + 
2 \sum_{k<j} \cos \theta_{kj} \sqrt{q(A=\alpha_j) q(B=\beta_i \vert A=\alpha_j)},
\end{equation}
where the additional parameter $\theta_{kj}$ are combination of phases of $\psi$ and eigenvectors of observables $A, B.$

We stress that, to get an additional (interference) term in FTP, we need not assume the linear space structure, 
as in quantum probability. It is sufficient to assume context-dependence of probability as implied by  {\bf PCOM 3}
(see author's works \cite{KHR4,PR77}), or in other words, the absence of jpd.

{\it Transformation of classical FTP into quantum FTP is one of the main roots of quantum superiority.} 
If the interference term is positive ({\it constructive interference of probabilities}), then 
\begin{equation}
\label{ADx}
q(B=\beta_i) - p(B=\beta_i) >0
\end{equation}
(the quantum and  classical probabilities are given by corresponding FTPs).

Thus an algorithm exploring the quantum probability calculus can essentially increase the probability of some event. We stress that this 
can happen only by exploring incompatible quantum observables. They are represented by noncommuting operators, i.e., $[A, B]\not=0.$
If observables are compatible, i.e., $[A, B] =0,$ then quantum FTP is reduced to classical FTP.  

The computational power of operating with incompatible observables is based on {\bf PCOM 5}: the possibility to extract the complete 
information about system's state with the aid of such observables. So, any event $E$ can be represented in the form $E=\{B=\beta_i\}$
for some quantum observable $B.$ 

\subsection{Quantum versus classical  inference}
\label{QV}

Let us remind that Bayes formula (\ref{AD2}) and FTP (\ref{AD3}) are the basic elements of classical probability update, the procedure know as probability inference. Quantum probability calculus represents a different procedure for probability inference. This new form 
of probability update leads to assigning to event $\{B=\beta_i\}$ probability  $q(B=\beta_i)$ that can be higher than probability 
$p(B=\beta_i)$ obtained via classical probability update. 

Quantum FTP (\ref{AD3}) can lead to increase of probability for some event.
However, this is just the formal explanation of quantum superiority. 
Yes, mathematics implies  (\ref{ADx}). What does it mean from the computational viewpoint?
Here we go back to sections \ref{CP}, \ref{QP} and point that the basic feature of the quantum probability calculus 
is the possibility to proceed without calculation of jpds. 

{\it This possibility 
to get rid of  calculations of jpds in combination with quantum conditioning via state's projection is the origin of probability increase, see (\ref{ADx}), 
implied by the quantum  probability calculus. This is the computational basis 
of supremacy of quantum algorithms. } 

\subsection{Algorithms based on  nonclassical  inferences}

Our previous considerations lead to the following hypothesis.

 {\it The essence of ``quantum superiority'' in the use of nonclassical  probability inference.} 

In connection with this hypothesis, the natural question arises: 

{\it Can one approach superiority over some classical 
probabilistic  algorithms by using nonclassical  probability inferences different from the quantum inference?} 

Formally, it seems to be possible and this is an interesting topic for studies in computer science. We shall make some remarks on this issue in the following sections. 

\section{Frequency realization of probability and its role for quantum computing}
\label{FR}
 
We  recall the basic law of classical probability theory, {\it the law of large numbers.} 
By this law the frequency of independent realizations of some event $E,$ in a long series of trials, $\nu_N = k_N(E)/N,$ where $k_N(E)$ is the number of occurrences of $E,$ approaches the probability of this event, 
\begin{equation}
\label{LLN}
p(E)= \lim_{N \to \infty}  \nu_N,
\end {equation} i.e., 
\begin{equation}
\label{LLN1}
p(E)\approx \nu_N
\end {equation}
for large $N.$ This law (which rigorous 
mathematical formalization was done by Kolmogorov and Khinchin) is the basis of experimental verification of probability.

Now, we make one remark on applicability of the law of large numbers in quantum probability theory. For any set of compatible quantum observables, it is possible  to construct a classical probability space, i.e., to use Kolmogorov probability. In particular, for a single quantum observable or  a vector of compatible quantum observables, we can use the frequency interpretation of probability.

All probabilistic algorithms are based on the frequency realization of probability, the law of large numbers. This procedure of generation of a single output of a probabilistic algorithm can be time consuming. Superiority condition (\ref{ADx})  is coupled to time interval $\Delta t$ needed for output's generation.

For quantum algorithms, we can try to couple  $\Delta t$ with {\it decoherence time} (section \ref{PP}) and show that output's generation is really rapid. But, by attempting to model quantum measurement as decoherence, we lose the solid ground of quantum theory and confront with its most fundamental problem, {\it the measurement problem.} 
    
\section{Role of measurement in quantum computing} 
\label{PP}

As is well known, quantum theory suffers of a bunch of unsolved foundational problems. The majority of quantum community, especially its young generation, follow Feynman's slogan: ``Shut up and calculate!'' During many years, this pragmatic approach was the only possible strategy: otherwise one would drown in the sea of foundational problems. However, soon or later one has to take foundations seriously,
as in the case of entanglement and its role in quantum computing. 

One of the most fundamental problems of quantum theory is the {\it measurement problem.} 
In the standard quantum formalism, measurement is described operationally with the aid of the projection postulate. An observable is represented  by the Hermitian operator 
\begin{equation}
\label{T1}
 A =\sum_x x\;  E^A(x),
\end{equation}
where $E^A(x)$ is the orthogonal projector on the subspace $H_{x}$ composed of eigenvectors with the eigenvalue $x$
(we consider only observables with discrete spectra). 
 The value $x$  is observed with the probability 
\begin{equation}
\label{T2a}
q^A(x)\equiv q(A=x; \rho_0) = \rm{Tr} \rho_0  E^A(x) 
\end{equation}
and the post-observation state with specified value $x$ is given by 
\begin{equation}
\label{T2}
 \rho^A_x =  \frac{E^A(x)  \rho_0  E^A(x)}{\rm{Tr} E^A(x) \;  \rho_0 \; E^A(x)}
\end{equation}
and without output specification by the state
\begin{equation}
\label{T3}
\rho^A =  \sum q^A(x)\;   \rho^A_x.
\end{equation}

If $A$-measurement with output $A=x$ is followed by the $B$-measurement with output $B=y,$ then quantum conditional probability 
$q(B=y\vert A=x)$ is defined simply  a the probability of  output $B=y$ for the state $\rho^A_x:$
\begin{equation}
\label{T2at}
q(B=y\vert A=x) \equiv q(B=y; \rho^A_x) = \rm{Tr} \rho^A_x  E^B(y), 
\end{equation}
where $E^B(y)$ is the spectral family of Hermitian operator $B.$

In particular, for pure initial state $\vert \psi_0 \rangle$ (with the density operator $\hat \rho_0= \vert \psi_0 \rangle \langle \psi_0\vert )$ 
the post-measurement state is always again the pure state: 
\begin{equation}
\label{T4}
\vert \psi^A_x \rangle =E^A(x) \vert \psi_0\rangle/\Vert E^A(x) \;\vert \psi_0 \rangle \Vert
\end{equation}
and 
\begin{equation}
\label{T2at}
q^A(x)= \Vert E^A(x) \;\vert \psi_0 \rangle \Vert^2.
\end{equation}
For the conditional probability, we have the following equality:
\begin{equation}
\label{T2att}
q(B=y\vert A=x) \equiv q(B=y;\psi^A_x) = \langle \psi^A_x\vert E^B(y) \vert \psi^A_x\rangle= 
\Vert E^B(y) E^A(x) \psi_0 \Vert^2/ \Vert E^A(x) \psi_0 \Vert^2. 
\end{equation}

The equality (\ref{T4}) expresses the L\"uders projection postulate. The projection-type of state transition makes the impression that this {\it transition is instantaneous.}
This instantaneous jump is in the striking contrast with continuous unitary evolution of a state. This situation was described 
by von Neumann as {\it the measurement problem.} This problem of mathematical unification of measurement and unitary evolution has not been 
resolved up to now.\footnote{We point to recent progress, see \cite{Theo}.} 
Appealing to the projection postulate, makes the impression  that the {\it measurement part of quantum 
computing can be accomplished instantaneously.} The presence of such an 
instantaneous operation mystifies quantum computing.

Of course, quantum measurement is not instantaneous, this is the process of complex interaction between a system $S$ and a measurement instrument $M.$
So, it takes time. In theory of {\it open quantum systems}, this process is described by an {\em indirect measurement model}
(see,e.g. \cite{Ozawa}) For  a Hilbert space $\cH,$ the state space of a system $S,$  and a Hermitian operator $A$ symbolically representing an observable on $S,$ 
this is a quadruple $(\cK,R,U,M_A)$ 
consisting of a Hilbert space $\cK,$  a density operator $R$, a unitary operator $U$ 
on $\cH\otimes\cK$, and a Hermitian operator $M_A$ on $\cK$.  By this measurement model $(\cK,R,U,M_A)$,  the Hilbert space $\cK$ describes the ``probe system'' $M$ (measurement instrument), the unitary operator  $U$ describes the time-evolution of the composite system $S+M$, the density operator  $R$ describes the initial state of the probe $M$, and the operator $M_A$ describes the meter observable in the probe $M$  to be measured directly just after the interaction. Then, the output probability distribution $q(x)$ 
is given by
\begin{equation}
\label{MA1}
q^A(x)= \rm{Tr} [(I\otimes E^{M_A}(x))U(\rho\otimes R)U^{\da}],
\end{equation}
where $E^{M_A}(x)$ is  the spectral projection of $M_A$ for the eigenvalue $x.$

The change of the state $\rh$ of the system $S$  caused by the measurement for the outcome $x$ is represented with the aid of 
the map $\cI_A(x)$ acting in the space of density operators defined as 
\begin{equation}
\label{MA1A}
 \rho^A_x\equiv \cI_A(x) \rho =\rm{Tr}_{\cK}[(I \otimes E^{M_A}(x)) U (\rho \otimes R) U^{\da}],
\end{equation}
where $\rm{Tr}_{\cK}$ is the partial trace over $\cK.$ The map  $x \to \cI_A(x)$ is a {\it quantum instrument.}

The unitary operator $U$ has the form $U=e^{-i \Delta t H},$ where $H$ is the Hamiltonian of $S+M.$ And 
the time interval $\Delta t$ gives the duration of measurement process.  This model demystifies (at least partially) 
the the measurement stage of quantum computing - it is not instantaneous,  it takes time. But, how much? $\Delta t=?$ It seems that this 
scheme cannot give the definite answer to this question. 
 
Nevertheless, we can try to get the answer to the question of the speed of accomplishment of a quantum measurement. As already 
emphasized, the quantum measurement problem has not been resolved. However, it is promising to apply the decoherence scheme. The initial step in this direction was done by Zurek \cite{Zurek}. However, Zurek tried to solve 
the genuine quantum measurement problem. And he was strongly criticized; his approach was not accepted as the real solution of this problem. In the series of papers (see the most recent work \cite{Asano} and references herein), I and my coauthors proposed a solution to so to say {\it a soft version of  the quantum measurement  problem.} We  modeled not the process of  generation of the concrete individual outputs of measurement, but the process of generation the classical probability distribution for these outputs.

Consider modeling of the quantum state dynamics in the process of measurement of an observable $A.$ We restrict our consideration to the case of observables which are mathematically represented by operators with nondegenerate spectra, i.e., 
\begin{equation}
\label{O}
A= \sum_x x \vert x\rangle \langle x \vert.
\end{equation}
To simplify consideration, we proceed under the assumption  that this dynamics is  Markovean. And it  is described {\it Gorini-Kossakowski-Sudarshan-Lindblad} (GKSL)
 equation \cite{ING}, 
\begin{equation}
\label{GKSL}
 \frac{d \rho}{d t}(t)= - [ H,   \rho(t)] +  \hat L \rho(t), \;  \rho(0)=  \rho_0,
\end{equation}
where $ H$ is Hamiltonian of $S$ and  $\hat L$ is a linear  operator acting in the space of  linear operators (a super-operator).  Commonly operator $ H$ represents the state dynamics in the absence of outer environment. The general situation is more complicated and
operator $\hat H$ can contain some environmental contribution. For ``natural'' systems, environments, and interactions  (encoded in operators 
$ H$ and $ L)$ the state $\rho(t)$ asymptotically approaches some steady state $\rho_A,$ the solution of the stationary equation:
\begin{equation}
\label{GKSLoo}
 [ H,  \rho_A] =   L  \rho_A .
\end{equation}
This state is considered as the post-measurement state. The state $\hat \rho_A$ is diagonal with respect to the $A$-basis
$(\vert x \rangle),$ i.e., it has the form 
\begin{equation}
\label{O1}
\rho_A= \sum_x \lambda_x \vert x\rangle \langle x \vert;
\end{equation}
the corresponding probability distribution is given by the equality
\begin{equation}
\label{O2}
q^A(x)= \lambda_x.
\end{equation}
Now, we remark that the GKSL-equation is a system of ordinary linear differential equations with  constant coefficients;
its solution is a linear combination of exponents $e^{(a+ib)t}.$ To have a steady state, we have to assume that $a<0.$ 
Thus the state $\rho(t)$ stabilizes {\it exponentially rapidly.} The main problem is that this is  teh scenarios foir the state and not measurement's output  stabilization. However, heuristically it is natural to assume the same  exponentially rapid dynamics 
for approaching of the eigenstates. (For the moment, this is just a speculation.)   

\section{Concluding remarks}

In the light of controversy generated by Google's claim on approaching superiority of quantum computer over classical digital computer,
it is important to reanalyze the foundational grounds of the possibility of such superiority.\footnote{We stress that we are interested 
quantum physical foundations (see, e.g., \cite{V1}, \cite{V2}-\cite{V4}), not in  foundations related to mathematics of computing.} Typically the latter was associated 
with entanglement (quantum nonlocality, action at a distance) or straightforwardly interpreted superposition. However, after the recent article \cite{KHR1} demonstrating locality of quantum mechanics, the mystical nonlocality  argument lost its power. 
The crucial difference between the classical (wave type) and genuine quantum superpositions was highlighted in \cite{KHR3}.
(Well, already Dirac emphasized this difference \cite{Dirac}, but nowadays not so many experts in quantum information read 
Dirac's works; see, however, \cite{Jaeger2}.) We have to  search for other possible roots of quantum supremacy. 

In this note (following \cite{KHR1,KHR4}), we coupled quantum computational supremacy with 
quantum complementarity, PCOM, see {\bf PCOM 1} -  {\bf PCOM 5}. The information interpretation of {\bf PCOM 3} (existence of complementary experimental contexts) and {\bf PCOM 5} (complementary observations  provide complete information 
about system's state) lead to their probabilistic reinterpretation. 

By analyzing the difference between classical and quantum probability calculi, we point to calculation of jpds (joint probability distributions) as the main source of consuming of computational 
resources in classical probabilistic algorithms. The quantum calculus can function without calculation of jpds, without  this (exponential) time consuming. This calculus opens the door to a new way of probability update, 
via quantum state update. This {\it non-Bayesian 
inference} is based on the modification of FTP expressing interference of probabilities. Constructive interference can essentially increase probability of some events, comparing with classical Bayesian inference. This {\it constructive interference is the main root 
of quantum computational superiority.} 

However, the recognition of the role of interference of probabilities does not complete our analysis. We elevated the role of the frequency realization of probability (mathematically the law of large numbers) and its coupling with  {\bf PCOM 4}, the basic distinguishing feature of quantum measurements - extraction of discrete phenomena. This issue highlights the role of linearity of measurement dynamics. The latter is closely coupled to the main unsolved problem of quantum theory, the measurement problem.

We can conclude that in principle algorithms based on the quantum probability calculus can demonstrate superiority over corresponding 
classical probabilistic algorithms, via saving computational resources by eliminating jpds-calculations and by exploring quantum inference, based on the constructive interference of probabilities. The crucial role is played by the discrete structure of quantum measurements and linearity of quantum state dynamics (as was speculated in section \ref{PP}).   

In principle, we can say that  quantum computational superiority is based the genuine quantum principle of superposition 
(as the linear space representation of  Bohr's PCOM) completed with (still hypothetical) linear dynamics of quantum measurements as 
well discreteness of outputs of observables and the  completeness assumption (see {\bf PCOM}). 

In Appendix 2, comparison of classical and quantum probability calculi  is transferred into comparison of classical and quantum 
logical structures used in classical and quantum algorithms.   

The final messages of this note is that technological research on quantum computing and generally quantum information technology has to be accomplished by intensive foundational studies (in the spirit of \cite{V1}, \cite{V2}-\cite{V4}). In particular, such studies have to be proportionally supported.  (By foundations I mean the real foundations, not just mathematics in complex Hilbert space.) For the moment, support of foundational studies is really miserable (comparing with support of quantum  technologies).  

\section*{Appendix 1: On quantum-like superiority}

Here we continue considerations of section \ref{QV}.

The information interpretation of some components of PCOM can make the impression that quantum-like computational superiority can be realized on systems which are not genuine quantum. The main candidates are classical optical devices.  The main obstacle is impossibility 
to generate discrete phenomena, the difference between classical and quantum superpositions (section \ref{SUP}). However,  
interference of probabilities, FTP with the interference term (\ref{AD3}), and hence approaching probability increasing  need not 
be based on the linear space representation of states of systems (see \cite{KHR4}). Thus to mimic quantum superiority  we need not be addicted to say classical optics. 

 However, it seems that the use of genuine quantum systems plays the fundamental role. It seems that {\bf PCOM 5} is crucial. 
Of course, we can mimic incompatibility, discrete clicks (with e.g., threshold type detectors \cite{D1,D2}), and even increase probability of some events through non-Bayesian inference based on FTP with interference term. It seems that we would not be able to guarantee 
that through measurements of incompatible observables we would extract the complete information about system's state. 

\section*{Appendix 2: Quantum versus classical logic}

The above quantum versus classical probability argument can be rephrased into the  form of {\it quantum versus classical logic argument.} We recall that in the Kolmogorov probability model events form a Boolean algebra. The quantum probability is based on {\it a partial Boolean algebra,} collection of Boolean algebras based on algebras of commuting Hermitian operators.

 A classical probabilistic algorithm (at least in the straightforward fashion)  can operate only in a Boolean algebra. It has to unify subalgberas of a partial Boolean algebra into the  complete  Boolean algebra. For this aim, such an algorithm has to construct all conjunctions and disjunctions of all events from subalgebras and iterate this process up to construction of the Boolean algebra. This process implies exponential computational time. 

Thus,  superiority of quantum computing is based on the use of quantum logic, instead of classical logic.
Since quantum logic can be considered as generalization of classical logic, it is natural to expect that quantum logic based reasoning 
is more powerful than classical reasoning. However, it is difficult to point to the concrete 
computational advantage of the use of quantum logic. The main difference between classical Boolean logic and quantum logic is violation of the distributivity axiom: {\it quantum conjunction of disjunction is not equal to disjunction of conjunctions} 
\begin{equation}
\label{ADx8}
 A \wedge (B \vee C)\not= (A \wedge B) \vee (A \wedge C).
\end{equation}
It is not clear how this can help to speed up calculations. As the first step, we have to analyze the structure of 
classical algorithms (with which the quantum algorithms compete) to understand the role of conjunction-disjunction 
distributivity; say, that this operation is used many times and its application leads to exponential computation time. 

As we know, quantum logic is just one of variety of nonclassical logical structures. Therefore, we can guess that some of them 
also can lead to essential speed up of algorithms.  However, we again have to point to the distinguishing feature of  
quantum logic, its linear representation. We guess that this is  crucial for speed up of calculations.

\end{document}